# Mathematical analysis of a two-strain disease model with amplification

Md Abdul Kuddus[1, 2, 5]*, Michael T. Meehan[1], Adeshina I. Adekunle[1], Lisa J. White[3, 4], Emma S. McBryde[1, 2]

[1]*Australian Institute of Tropical Health and Medicine, James Cook University, Townsville, QLD*
[2]*College of Medicine and Dentistry, James Cook University, Townsville, QLD*
[3]*Mahidol Oxford Tropical Medicine Research Unit (MORU), Bangkok, Thailand*
[4]*Centre for Tropical Medicine and Global Health, Nuffield Department of Clinical Medicine, University of Oxford, Oxford, United Kingdom*
[5]*Department of Mathematics, University of Rajshahi, Rajshahi-6205, Bangladesh*

## ABSTRACT

We investigate a two-strain disease model with amplification to simulate the prevalence of drug-susceptible (s) and drug-resistant (m) disease strains. We model the emergence of drug resistance as a consequence of inadequate treatment, i.e. amplification. In this case, individuals infected with the drug-susceptible strain acquire drug-resistant infection such that the strains are coupled. We perform a dynamical analysis of the resulting system and find that the model contains three equilibrium points: a disease-free equilibrium; a mono-existent disease-endemic equilibrium with respect to the drug-resistant strain; and a co-existent disease-endemic equilibrium where both the drug-susceptible and drug-resistant strains persist. We found two basic reproduction numbers: one associated with the drug-susceptible strain $(R_{0s})$; the other with the drug-resistant strain $(R_{0m})$, and showed that at least one of the strains can spread in a population if $\max[R_{0s}, R_{0m}] > 1$ (epidemic), but are maintained in that population without the need for external inputs when the effective reproduction number is equal to 1. Furthermore, we also showed that if $R_{0m} > \max[R_{0s}, 1]$, the drug-susceptible strain dies out but the drug-resistant strain persists in the population; however if $R_{0s} > \max[R_{0m}, 1]$, then both the drug-susceptible and drug-resistant strains persist in the population. We conducted a local stability analysis of the system equilibrium points using the Routh-Hurwitz conditions and a global stability analysis using appropriate Lyapunov functions. Sensitivity analysis was used to identify the most important model parameters through the partial rank correlation coefficient (PRCC) method. We found that the contact rate of both strains had the largest influence on prevalence. We also investigated the impact of amplification and treatment/recovery rates of both strains on the equilibrium prevalence of infection; results suggest that poor quality treatment/recovery (drugs) make coexistence more likely but increase the relative abundance of resistant infections.

## 1 Introduction

Many pathogens have several circulating strains. The presence of more than one strain of the pathogen is mostly due to incorrect treatment, poor adherence, malabsorption, and treatment with antibodies or antiviral drugs leading to the acquisition of resistance, i.e. amplification [1-3]. The growing threat of several strains including a drug-resistant strain presents a significant challenge throughout the world, particularly in developing countries and those with lower socio-economic status [4]. Once subsequent transmission of drug-resistant strains has emerged in a population, these strains may also contribute to the disease burden (in addition to amplification) [5]. Thus, one of the major challenges in preventing the spread of infectious diseases is to control the genetic variations of pathogens through proper treatment regimens [6, 7]. Recent studies [8-12] have shown that drug-resistant strains can possess higher virulence to transmit disease than drug-susceptible strains, and those individuals infected with a drug-resistant strain have the highest mortality rate, e.g. tuberculosis and HIV.

To examine the great threat posed by genetic variations of pathogens, we present a two-strain (drug-susceptible, and drug-resistant) SIR epidemic model with coupled infectious compartments and use it to investigate the emergence and spread of mutated strains of infectious diseases. We consider the possibility that an individual's position changes from drug-susceptible at initial presentation to resistant at follow-up. This is the mode by which drug resistance first emerges in a population and is designed to reproduce the phenotypic phenomenon of amplification. The model can be applied to investigate the co-existent or competitive exclusive phenomena among the strains.

Explicitly, in this paper we perform a rigorous analytical and numerical analysis of the proposed two-strain model properties and solutions from both the mathematical and biological viewpoints. For each, we used the next generation matrix method to determine analytic expressions of the basic reproduction numbers of the drug-susceptible and drug-resistant strains and found that these are important determinants for regulating system dynamics. With a focus on the early- and late-time behavior of the system, we outline the required conditions for disease fade-out, infection mono-existence, and co-existence.

To supplement and validate the analytic analysis, we use numerical techniques to solve the model equations and explore the epidemic trajectory for a range of possible parameter values and initial conditions. The local stability of three system equilibria is examined using the Routh-Hurwitz conditions and the global stability of the disease-free equilibrium and mono-existent disease-endemic equilibrium is examined using appropriate Lyapunov functions. Following this, we perform a sensitivity analysis to investigate the model parameters that have the greatest influence on disease prevalence.

The remainder of this paper is constructed as follows: in section 2 we present the two-strain SIR model with differential infectivity and amplification, and verify the boundedness and positivity of solutions as well as the existence of several equilibria. Local and global stability analyses of the equilibria are presented in section 3. In section 4 we discuss a sensitivity analysis of the model outputs. We then provide numerical simulations to support analytic results in section 5. Finally, in section 6, we provide a summary of our outcomes, discuss their importance for public health policy and propose guidelines for future effort.

## 2. Model description and analysis

**Model equations:**

We developed a transmission dynamic two-strain SIR model for drug-susceptible and drug-resistant cases, and divided the total population into four subclasses: $S$ — susceptible individuals; $I_s$ — individuals infected with the drug-susceptible strain; $I_m$ — individuals infected with the drug-resistant strain; and $R$ — recovered individuals, who are assumed to have immunity against both strains. Thus the total population number $N(t)$ at time t is

$$N(t) = S(t) + I_s(t) + I_m(t) + R(t). \tag{1}$$

We also introduced the following parameters: $\Lambda$ — constant recruitment rate into the susceptible class through birth or immigration; $\mu$ — natural death rate; $\beta_s$ ($\beta_m$) — effective contact rate of individuals with drug-susceptible (drug-resistant) infection; $\omega_s$ ($\omega_m$) — treatment/recovery rate for drug-susceptible (drug-resistant) infected individuals; $\phi_s$ ($\phi_m$) — disease related death rate for drug-susceptible (drug-resistant) infected individuals; $\rho$ — proportion of individuals who amplify from the drug-susceptible strain to the drug-resistant strain during treatment/recovery. We assumed the proportion of individuals who amplify — due to incomplete treatment or lack of compliance in the use of first-line drugs — move directly from the drug-susceptible compartment $I_s$ into the drug-resistant compartment $I_m$. The model structure is illustrated in Fig. 1.

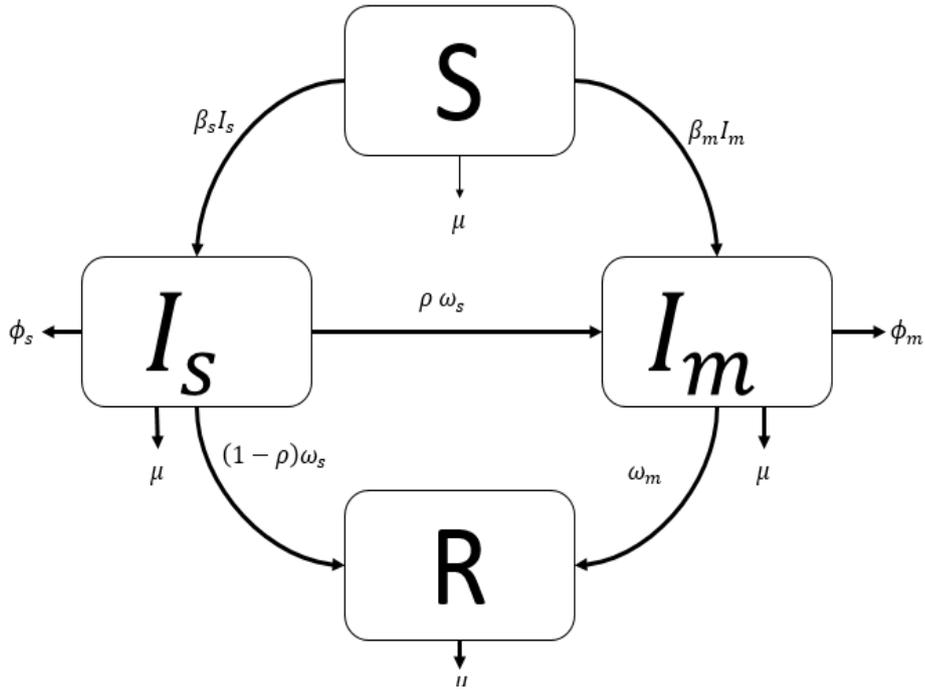

Fig. 1. Flow chart of the two-strain SIR model showing the four infection states, and the transition rates in and out of each state (not shown: the constant recruitment rate $\Lambda$ into the susceptible compartment S). Subscripts s and m denote drug-susceptible and drug-resistant quantities respectively.

From the aforementioned, the populations in each disease state are determined by the following system of nonlinear ordinary differential equations:

$$\dot{S} = \Lambda - \mu S - \beta_s I_s S - \beta_m I_m S, \tag{2}$$

$$\dot{I}_s = \beta_s I_s S - (\omega_s + \phi_s + \mu) I_s, \tag{3}$$

$$\dot{I}_m = \rho \omega_s I_s + \beta_m I_m S - (\omega_m + \phi_m + \mu) I_m, \tag{4}$$

$$\dot{R} = (1-\rho)\omega_s I_s + \omega_m I_m - \mu R. \tag{5}$$

Given non-negative initial conditions for the system above, it is straightforward to show that each of the state variables remain non-negative for all $t > 0$. Moreover, summing equations (2)–(5) we find that the size of the total population, $N(t)$ satisfies

$$\dot{N}(t) \leq \Lambda - \mu N.$$

Integrating this equation we find

$$N(t) \leq \frac{\Lambda}{\mu} + N(0)e^{-\mu t}.$$

This shows that the total population size $N(t)$ is bounded in this case and that it naturally follows that each of the compartment states S, I, etc. are also bounded.

Note that equations (2)-(4) are independent of the size of the recovered population $R(t)$; therefore, if we only wish to track disease incidence and prevalence, we can focus our attention on the following reduced system (6)–(8):

$$\dot{S} = \Lambda - \mu S - \beta_s I_s S - \beta_m I_m S, \tag{6}$$

$$\dot{I}_s = (\beta_s S - \chi_s) I_s, \tag{7}$$

$$\dot{I}_m = \rho \omega_s I_s + \beta_m I_m S - \chi_m I_m, \tag{8}$$

where, $\chi_s = \omega_s + \phi_s + \mu$ and $\chi_m = \omega_m + \phi_m + \mu$ represent the total removal rates from the respective infectious compartments.

Given the positivity and boundedness of the system solutions, we find that the feasible region for equations (6) – (8) is given by

$$D = \left\{ (S, \ I_s, \ I_m) \in \mathbb{R}_+^3 : S + I_s + I_m \leq \frac{\Lambda}{\mu} \right\} \tag{c}$$

where D is positively invariant. Therefore, in this study we consider the system of equations (6) – (8) in the set D.

## 2.1 Basic reproduction number

Here we estimate the basic reproduction number of the model (6)-(8). In an epidemic model, the basic reproduction number is an important quantity that reflects the expected number of secondary cases created by a single infectious case introduced into a totally susceptible population. If the basic reproduction number is greater than one, the number of infected individuals grows and the infection will show persistent behavior. Conversely, if the basic reproduction number is less than one, the number of infective individuals typically tends to zero [13-15]. Here we use the next generation matrix technique to estimate the basic reproduction number(s) of our system.

The reduced model (6)-(8) has two infected states, $I_s$ and $I_m$, and one uninfected state, S. At the infection-free steady state $I_s^0 = I_m^0 = 0$, hence, from (6), $S^0 = \frac{\Lambda}{\mu}$. When we linearize the system about the infection-free equilibrium equations $(3) - (4)$ are closed, and we have

$$\dot{I}_s = (\beta_s S^0 - \chi_s) I_s, \tag{9}$$

$$\dot{I}_m = \rho \omega_s I_s + \beta_m I_m S^0 - \chi_m I_m. \tag{10}$$

Here, the ODEs (9) and (10) are referred to as the (linearized) infection subsystem, as they only describe the production of newly infected individuals and changes in the states of already infected individuals.

By setting $\mathbf{x}^T = (I_s, \ I_m)^T$, where T denotes transpose, the infection subsystem can be written in the following form:

$$\dot{\mathbf{x}} = (T + \Sigma) \mathbf{x}. \tag{11}$$

The matrix $T$ corresponds to transmission (arrival of susceptible individuals into the infected compartments $I_s$ and $I_m$) and the matrix $\Sigma$ to transitions (i.e. recovery, amplification and death).

For the subsystem $(9) - (10)$ we obtain

$$T = \begin{pmatrix} \beta_s S^0 & 0 \\ 0 & \beta_m S^0 \end{pmatrix} \text{ and } \Sigma = \begin{pmatrix} -\chi_s & 0 \\ \rho\omega_s & -\chi_m \end{pmatrix}.$$

The next generation matrix, K, is given by [13]

$$K = -T\Sigma^{-1} = \begin{pmatrix} \frac{S^0 \beta_s}{\chi_s} & 0 \\ \frac{S^0 \beta_m \omega_s \rho}{\chi_s \chi_m} & \frac{S^0 \beta_m}{\chi_m} \end{pmatrix}.$$

The dominant eigenvalues of K are the basic reproduction numbers for the drug-susceptible and drug-resistant strains; they represent the average number of new infections from each strain produced by one infected individual. Hence, the basic reproduction numbers for the drug-susceptible and drug-resistant strains are:

$$R_{0s} = \frac{S^0 \beta_s}{\chi_s} = \frac{\Lambda \beta_s}{\mu \chi_s}, \tag{a}$$

and

$$R_{0m} = \frac{S^0 \beta_m}{\chi_m} = \frac{\Lambda \beta_m}{\mu \chi_m}. \tag{b}$$

Interestingly we find that the basic reproduction numbers $R_{0s}$ and $R_{0m}$ are purely a function of the epidemiological parameters of the drug-susceptible and drug-resistant strains respectively, i.e. both are independent of the amplification rate $\rho$ [16].

Given these expressions (a)-(b) for the basic reproduction numbers of each strain, we can now investigate the relationship between the fitness cost, c, exacted on the transmissibility of drug-resistant strain m, and its resistance to treatment, $\epsilon$, on the relative fitness of strains m and s.

If we assume that both $R_{0s}$ and $R_{0m}$ are greater than 1, then the condition for resistant infections to outcompete sensitive infections is given by,

$$R_{0m} > R_{0s}.$$

Substituting the formulae (a)-(b) for the basic reproduction numbers gives:

$$\frac{\Lambda \beta_m}{\omega_m + \phi_m + \mu} > \frac{\Lambda \beta_s}{\omega_s + \phi_s + \mu}.$$

If we consider that resistance exacts a fitness cost, c, on the transmissibility of the drug-resistant strain, it follows that

$$\beta_m = (1-c)\beta_s.$$

Further, if we assume that strain m has a level of resistance, $\epsilon$, to treatment we have

$$\omega_m = (1-\epsilon)\omega_s.$$

Lastly, assuming that $\mu \approx 0$ (since it is very slow compared to the other rates) and that $\phi_m \approx \phi_s$ yields the condition

$$\frac{\Lambda(1-c)\beta_s}{(1-\epsilon)\omega_s + \phi_s} > \frac{\Lambda\beta_s}{\omega_s + \phi_s}$$

which we can rearrange to obtain

$$\epsilon > \frac{c(\omega_s + \phi_s)}{\omega_s}.$$

The above relation shows that the resistant strain can outcompete the susceptible strain if the resistance level $\epsilon$ is high (which may be the case for drug-resistant individuals) and/or the death rate due to infection is high, which may occur as a result of poor medical care. Alternatively, the resistant strain will be fitter than the susceptible one if the fitness cost c is sufficiently low.

## 2.2 System properties

**Table 1:** Description and estimation of model parameters

| Parameters | Description | Estimated value | References |
|---|---|---|---|
| $\Lambda$ | Recruitment rate in to the population | 1 | [17] |
| $\mu$ | Death rate | $\frac{1}{70}$ per year | [17] |
| $\beta_s$ | Contact/transmission rate for drug susceptible population | variable | - - |
| $\beta_m$ | Contact/transmission rate for drug resistant population | variable | - - |
| $\omega_s$ | Recovery rate for drug susceptible population | 0.2906 per year | [18] |
| $\omega_m$ | Recovery rate for drug resistant population | 0.1453 per year | Assume |
| $\rho$ | Proportion of amplification | 0.035 | [19] |
| $\varphi_s$ | Disease related death rate for drug susceptible population | 0.37 over 3 year | [19] |
| $\varphi_m$ | Disease related death rate for drug-resistant population | 0.37 over 3 year | [19] |

### 2.2.2 Existence of equilibria

Clearly, equations (6) – (8) always have a disease-free equilibrium

$$E^0 = (S^0,\ I_s^0,\ I_m^0) = \left(\frac{\Lambda}{\mu},\ 0,\ 0\right).$$

From equation (6)-(8) we can also derive the mono-existent endemic equilibrium point $E^1 = (S^1, 0, I_m^1)$ at which the drug-resistant strain persists and the drug-susceptible strain dies out:

$$S^1 = \frac{S^0}{R_{0m}} = \frac{\Lambda}{\mu R_{0m}},$$

$$I_s^1 = 0,$$

$$I_m^1 = \frac{\mu(R_{0m}-1)}{\beta_m}. \tag{13}$$

From (13) we see that $E^1 = (S^1, 0, I_m^1) \in D$ if, and only if $R_{0m} \geq 1$.

Next, the co-existent endemic equilibrium of the system is examined. If $E^2 = (S^2, I_s^2, I_m^2)$ is any co-existent endemic equilibrium, from equations (6)—(8), we obtain

$$S^2 = \frac{\Lambda}{\mu R_{0s}} = \frac{S^0}{R_{0s}},$$

$$I_s^2 = \frac{\mu(R_{0s}-1)}{\beta_s} \Psi, \text{ where } \Psi = \left(1 + \frac{\rho\omega_s R_{0s}\beta_m}{\beta_s\chi_m(R_{0s}-R_{0m})}\right)^{-1} = \left(1 + \frac{\rho\omega_s R_{0m}}{\chi_s(R_{0s}-R_{0m})}\right)^{-1} > 0 \text{ if } R_{0s} > R_{0m},$$

$$I_m^2 = \frac{\rho R_{0s}\omega_s\mu(R_{0s}-1)}{\beta_s\chi_m(R_{0s}-R_{0m})+\rho R_{0s}\omega_s\beta_m}, \tag{14}$$

$$= \frac{\rho\omega_s R_{0s}}{\chi_m(R_{0s}-R_{0m})} \frac{\mu(R_{0s}-1)}{\beta_s}\Psi.$$

Equation (14) shows that if $R_{0s} > \max[R_{0m}, 1]$ then the co-existent endemic equilibrium $E^2 = (S^2, I_s^2, I_m^2) \in D$.

## 3. Stability analysis

Since equations (2)–(4) are independent of equation (5), we can focus our attention on equations (6)-(8) to study the persistence of the infection. To investigate stability of the equilibria of equations (6)—(8), the following results are established:

### 3.1 Infection-free equilibrium

**Lemma 1:** If $R_0 = \max[R_{0s}, R_{0m}] < 1$, the disease free equilibrium $E^0$ of (6)—(8) is locally and globally asymptotically stable; if, however, $R_0 = \max[R_{0s}, R_{0m}] > 1$, $E^0$ is unstable.

**Proof:** We consider the Jacobian of the system (6)—(8) which is given by

$$J = \begin{pmatrix} -\beta_s I_s - \beta_m I_m - \mu & -\beta_s S & -\beta_m S \\ \beta_s I_s & \beta_s S - \chi_s & 0 \\ \beta_m I_m & \rho\omega_s & \beta_m S - \chi_m \end{pmatrix}.$$

At the infection-free equilibrium point, $E^0$, this reduces to

$$J_0 = \begin{pmatrix} -\mu & -\beta_s S^0 & -\beta_m S^0 \\ 0 & \chi_s(R_{0s}-1) & 0 \\ 0 & \rho\omega_s & \chi_m(R_{0m}-1) \end{pmatrix}.$$

The structure of $J_0$ allows us to immediately read off the 3 eigenvalues, $\lambda_i$, as

$$\lambda_1 = -\mu, \lambda_2 = \chi_s(R_{0s}-1), \text{ and } \lambda_3 = \chi_m(R_{0m}-1). \tag{15}$$

It is easy to verify that all the roots of the characteristic equation (15) have negative real parts for $R_{0s} < 1$ and $R_{0m} < 1$. Hence, the disease free equilibrium $E^0$ of (15) is locally asymptotically stable for $R_{0s} < 1$ and $R_{0m} < 1$. If $R_{0s} > 1$ or $R_{0m} > 1$, at least one of the roots of the characteristic equation (15) has a positive real part. Hence, in this case, $E^0$ is unstable.

Now the global stability of the disease free equilibrium $E^0$ for $R_{0s} < 1$ and $R_{0m} < 1$ can be investigated. In fact, from equation (7), we have

$$\dot{I}_s = (\beta_s S - \chi_s) I_s.$$

Integrating gives

$$I_s(t) = I_s(0) e^{\int_0^t \beta_s S(\tau) d\tau - \chi_s t} \tag{16}$$

for all $t \geq 0$.

Substituting in the condition $S(t) \leq \frac{\Lambda}{\mu} = S^0$, which follows immediately from the definition of D (equation (c)), we obtain

$$I_s(t) \leq I_s(0) e^{\int_0^t \beta_s \left(\frac{\Lambda}{\mu} + S(0) e^{-\mu \tau}\right) d\tau - \chi_s t},$$

$$\leq I_s(0) e^{(\beta_s S^0 - \chi_s) t}$$

$$\leq I_s(0) e^{\chi_s (R_{0s} - 1) t}$$

It follows then that if $R_{0s} < 1$ we have $I_s(t) \to 0$ as $t \to \infty$.

Hence the hyperplane $I_s = 0$ attracts all solutions of (6)–(8) originating in D whenever $R_{0s} < 1$.

Since $I_s(t) \to 0$ as $t \to \infty$ for $R_{0s} < 1$, it follows that $\rho \, \omega_s I_s(t) \to 0$, such that equation (8) reduces to

$$\dot{I}_m = \beta_m I_m S - \chi_m I_m.$$

Following the same strategy for $I_m$ as we used above for $I_s$ yields

$$I_m(t) \leq I_m(0) e^{\chi_m (R_{0m} - 1) t}.$$

It follows then that if $R_{0m} < 1$ that $I_m(t) \to 0$ as $t \to \infty$ and the hyperplane $I_m = 0$ attracts all solutions of (6)–(8) originating in D. Moreover, it is straightforward to show that if $I_s \to 0$, and $I_m \to 0$, then $S \to S^0$. Therefore $E^0$ is globally asymptotically stable when $R_0 = \max[R_{0s}, R_{0m}] < 1$.

### 3.2 Mono-existent endemic equilibrium

**Lemma 2:** If the boundary equilibrium $E^1 = (S^1, 0, I_m^1)$ of the equations (6)–(8) exists and $R_{0m} > \max[1, R_{0s}]$, $E^1$ is locally and globally asymptotically stable.

**Proof:** We consider the Jacobian of the system (6)–(8) at the mono-existent endemic equilibrium point $E^1$ which is given by

$$J_1 = \begin{pmatrix} -\beta_m I_m^1 - \mu & -\beta_s S^1 & -\beta_m S^1 \\ 0 & -\dfrac{\chi_s(R_{0m} - R_{0s})}{R_{0m}} & 0 \\ \beta_m I_m^1 & \rho\omega_s & 0 \end{pmatrix}.$$

The structure of $J_1$ allows us to immediately read off the first eigenvalue, $\lambda_1 = -\chi_s \dfrac{(R_{0m} - R_{0s})}{R_{0m}}$, and the remaining eigenvalues we can calculate from the following expression

$$(\lambda^2 + a_1\lambda + a_2) = 0 \tag{17}$$

where,

$a_1 = \beta_m I_m^1 + \mu = \mu R_{0m}$,

$a_2 = \beta_m^2 I_m^1 S^1 = \mu\chi_m(R_{0m} - 1)$.

For local stability we must ensure that the Routh-Hurwitz criteria [20] are satisfied:

$a_1 > 0$, which is true;

$a_2 > 0$, which holds whenever $R_{0m} > 1$.

Finally, the remaining root of the Jacobian $J_1$ is $\lambda_1 = -\chi_s \dfrac{(R_{0m} - R_{0s})}{R_{0m}} < 0$ for $R_{0m} > R_{0s}$. Thus, by the Routh-Hurwitz criteria, the boundary equilibrium $E^1$ is locally asymptotically stable whenever $R_{0m} > \max[1, R_{0s}]$. Conversely, for $E^1 \in D$, it is unstable when $R_{0m} < R_{0s}$.

Now we prove $E^1$ is globally asymptotically stable if $R_{0m} > \max[1, R_{0s}]$. Considering equation (7) and (8), we get

$$\dot{I}_s = (\beta_s S - \chi_s)I_s, \tag{18}$$

$$\dot{I}_m = \rho\omega_s I_s + \beta_m I_m S - \chi_m I_m. \tag{19}$$

Following [21] first, we divide equation (18) and (19) through by $I_s$ and $I_m$ respectively to obtain

$$\frac{d\log I_s}{dt} = \beta_s S - \chi_s \tag{20}$$

$$\frac{d\log I_m}{dt} = \beta_m S - \chi_m + \rho\omega_s \frac{I_s}{I_m} \tag{21}$$

Rearranging equations (20) and (21) to solve for S we get

$$S = \frac{1}{\beta_s}\frac{d\log I_s}{dt} + \frac{\chi_s}{\beta_s} = \frac{1}{\beta_m}\frac{d\log I_m}{dt} + \frac{\chi_m}{\beta_m} - \frac{\rho\omega_s}{\beta_m}\frac{I_s}{I_m}. \tag{22}$$

From (22), we obtain the following inequality

$$\frac{1}{\beta_s}\frac{d\log I_s}{dt} + \frac{\chi_s}{\beta_s} \leq \frac{1}{\beta_m}\frac{d\log I_m}{dt} + \frac{\chi_m}{\beta_m}$$

Integrating both sides gives

$$\left(\frac{I_s(t)}{I_s(0)}\right)^{\frac{1}{\beta_s}} e^{\frac{\chi_s}{\beta_s}t} \leq \left(\frac{I_m(t)}{I_m(0)}\right)^{\frac{1}{\beta_m}} e^{\frac{\chi_m}{\beta_m}t}$$

which we can rearrange to obtain

$$\left(\frac{I_s(t)}{I_s(0)}\right)^{\frac{1}{\beta_s}} \leq \left(\frac{I_m(t)}{I_m(0)}\right)^{\frac{1}{\beta_m}} e^{\left(\frac{\chi_m}{\beta_m}-\frac{\chi_s}{\beta_s}\right)t}.$$

Using equations (a) and (b) for the basic reproduction numbers, we get

$$\left(\frac{I_s(t)}{I_s(0)}\right)^{\frac{1}{\beta_s}} \leq \left(\frac{I_m(t)}{I_m(0)}\right)^{\frac{1}{\beta_m}} e^{S^0\left(\frac{1}{R_{0m}}-\frac{1}{R_{0s}}\right)t}.$$

Since both $I_s(t)$ and $I_m(t)$ are bounded, taking the limit as $t \to \infty$ we find

$$\lim_{t\to\infty}\left(\frac{I_s(t)}{I_s(0)}\right)^{\frac{1}{\beta_s}} \leq \lim_{t\to\infty}\left(\frac{I_m(t)}{I_m(0)}\right)^{\frac{1}{\beta_m}} e^{S^0\left(\frac{1}{R_{0m}}-\frac{1}{R_{0s}}\right)t} \to 0 \text{ for } R_{0m} > R_{0s}.$$

Hence the hyperplane $I_s = 0$ attracts all solutions of (6) – (8) when $R_{0m} > R_{0s}$.

We now show the endemic equilibrium $E^1$ is globally asymptotically stable on the hyperplane $I_s = 0$ by constructing the following Lyapunov function [22]:

Let

$$V_1 = S - S^1 \ln S + I_m - I_m^1 \ln I_m + C$$

where

$$C = -(S^1 - S^1 \ln S^1 + I_m^1 - I_m^1 \ln I_m^1).$$

Taking the derivative of $V_1(t)$ along system trajectories yields

$$\dot{V}_1 = \left(1 - \frac{S^1}{S}\right)\dot{S} + \left(1 - \frac{I_m^1}{I_m}\right)\dot{I}_m,$$
$$= \left(1 - \frac{S^1}{S}\right)(\Lambda - \mu S - \beta_m I_m S) + \left(1 - \frac{I_m^1}{I_m}\right)(\beta_m I_m S - \chi_m I_m)$$
$$= \Lambda - \mu S - \beta_m I_m S - \Lambda\frac{S^1}{S} + \mu S^1 + \beta_m I_m S^1 + \beta_m I_m S - \chi_m I_m - \beta_m I_m^1 S + \chi_m I_m^1.$$

First, we substitute in the identity

$$\Lambda = \mu S^1 + \beta_m I_m^1 S^1$$

to obtain

$$\dot{V}_1 = \mu S^1 + \beta_m I_m^1 S^1 - \mu S - \mu S^1\frac{S^1}{S} - \beta_m I_m^1 S^1\frac{S^1}{S} + \mu S^1 + \beta_m I_m S^1 - \chi_m I_m - \beta_m I_m^1 S + \chi_m I_m^1$$
$$= \mu S^1\left(2 - \frac{S}{S^1} - \frac{S^1}{S}\right) + \beta_m I_m^1 S^1 - \beta_m I_m^1 S^1\frac{S^1}{S} + \beta_m I_m S^1 - \chi_m I_m - \beta_m I_m^1 S + \chi_m I_m^1.$$

We can simplify this expression further by substituting in the identity

$$\beta_m S^1 = \chi_m$$

to get

$$\dot{V}_1 = \mu S^1\left(2 - \frac{S}{S^1} - \frac{S^1}{S}\right) + \chi_m I_m^1 - \chi_m I_m^1\frac{S^1}{S} + \chi_m I_m - \chi_m I_m - \chi_m I_m^1\frac{S}{S^1} + \chi_m I_m^1$$

$$= \mu S^1 \left(2 - \frac{S}{S^1} - \frac{S^1}{S}\right) + \chi_m I_m^1 \left(2 - \frac{S}{S^1} - \frac{S^1}{S}\right)$$

$$= (\mu S^1 + \chi_m I_m^1)\left(2 - \frac{S}{S^1} - \frac{S^1}{S}\right).$$

Since the arithmetic mean is greater than or equal to the geometric mean, we obtain

$\dot{V}_1 \leq 0.$

Therefore, the mono-existent endemic equilibrium $E^1$ is globally asymptotically stable if $R_{0m} > 1$.

### 3.3 Co-existent endemic equilibrium

We now show the stability analysis of the co-existent endemic equilibrium $E^2 = (S^2, I_s^2, I_m^2)$.

**Lemma 3:** If the endemic equilibrium $E^2 = (S^2, I_s^2, I_m^2)$ of the equations (6)—(8) exists, $E^2$ is locally asymptotically stable.

**Proof:** We consider the Jacobian of the system (6)—(8) at the co-existent endemic equilibrium point $E^2$ which is given by

$$J_2 = \begin{pmatrix} -\beta_s I_s^2 - \beta_m I_m^2 - \mu & -\beta_s S^2 & -\beta_m S^2 \\ \beta_s I_s^2 & \beta_s S^2 - \chi_s & 0 \\ \beta_m I_m^2 & \rho \omega_s & \beta_m S^2 - \chi_m \end{pmatrix}.$$

Now

$-\beta_s I_s^2 - \beta_m I_m^2 - \mu = -\mu R_{0s},$

$-\beta_s S^2 = -\frac{R_{0s} \chi_s}{S^0} \frac{S^0}{R_{0s}} = -\chi_s,$

$-\beta_m S^2 = -\frac{R_{0m} \chi_m}{S^0} \frac{S^0}{R_{0s}} = -\chi_m \frac{R_{0m}}{R_{0s}},$

$\beta_s I_s^2 = \beta_s \frac{\mu(R_{0s}-1)}{\beta_s} \Psi = \mu(R_{0s} - 1)\Psi,$

$\beta_m I_m^2 = \frac{\rho \omega_s}{\chi_s} \frac{R_{0m}}{(R_{0s} - R_{0m})} \mu(R_{0s} - 1)\Psi,$

$\beta_s S^2 - \chi_s = \chi_s \left(\frac{\beta_s S^0}{R_{0s}\chi_s} - 1\right) = \chi_s \left(\frac{R_{0s}}{R_{0s}} - 1\right) = \chi_s(1 - 1) = 0,$

$\beta_m S^2 - \chi_m = \chi_m \left(\frac{\beta_m S^0}{\chi_m R_{0s}} - 1\right) = \frac{\chi_m}{R_{0s}}(R_{0m} - R_{0s}).$

Given the identities above, the matrix $J_2$ will be in the following form,

$$J_2 = \begin{pmatrix} -\mu R_{0s} & -\chi_s & -\chi_m \frac{R_{0m}}{R_{0s}} \\ \mu(R_{0s} - 1)\Psi & 0 & 0 \\ \frac{\rho \omega_s}{\chi_s} \frac{R_{0m}}{(R_{0s} - R_{0m})} \mu(R_{0s} - 1)\Psi & \rho \omega_s & \chi_m \frac{(R_{0m} - R_{0s})}{R_{0s}} \end{pmatrix}.$$

To determine the stability of this matrix we use the Routh-Hurwitz criteria. Specifically, all of the roots of the characteristic polynomial associated with a three by three matrix $J_2$ are negative if $A_1 > 0, A_2 > 0, A_3 > 0$, and $A_1 A_2 > A_3$, where $A_1 = -\text{trace}(J_2)$, $A_2$ represents the sum of the two by two principal minors of $J_2$ and $A_3 = -\det(J_2)$.

**Condition 1:** For the matrix $J_2$, we have

$A_1 = -\text{trace}(J_2) > 0$,

$\Rightarrow \mu R_{0s} + \frac{\chi_m (R_{0s} - R_{0m})}{R_{0s}} > 0$, which is true if $R_{0s} > R_{0m}$.

**Condition 2:**

$$A_2 = \begin{vmatrix} 0 & 0 \\ \rho \omega_s & \frac{\chi_m (R_{0m} - R_{0s})}{R_{0s}} \end{vmatrix} + \begin{vmatrix} -\mu R_{0s} & -\chi_m \frac{R_{0m}}{R_{0s}} \\ \frac{\rho \omega_s}{\chi_s} \frac{R_{0m}}{(R_{0s} - R_{0m})} \mu(R_{0s} - 1)\Psi & \chi_m \frac{(R_{0m} - R_{0s})}{R_{0s}} \end{vmatrix}$$

$$+ \begin{vmatrix} -\mu R_{0s} & -\chi_s \\ \mu(R_{0s} - 1)\Psi & 0 \end{vmatrix} > 0,$$

$\Rightarrow \mu \chi_m (R_{0s} - R_{0m}) + \frac{\rho \omega_s}{\chi_s} \frac{\chi_m R_{0m}^2}{R_{0s}} \frac{\mu (R_{0s} - 1)}{(R_{0s} - R_{0m})} \Psi + \mu \chi_s (R_{0s} - 1)\Psi > 0$,

$\Rightarrow \mu \chi_m (R_{0s} - R_{0m}) + \mu(R_{0s} - 1)\Psi \left[ \frac{\rho \omega_s}{\chi_s} \frac{\chi_m R_{0m}^2}{R_{0s} (R_{0s} - R_{0m})} + \chi_s \right] > 0.$

Recalling the definition of $\Psi$ (equation (14)), which is positive for $R_{0s} > R_{0m}$, we see that this condition is satisfied whenever $R_{0s} > 1$ and $R_{0s} > R_{0m}$.

**Condition 3:**

$A_3 = \det(J_2) < 0$,

$\Rightarrow -\mu (R_{0s} - 1)\Psi \begin{vmatrix} -\chi_s & \chi_m \frac{R_{0m}}{R_{0s}} \\ \rho \omega_s & \chi_m \frac{(R_{0m} - R_{0s})}{R_{0s}} \end{vmatrix} < 0$,

$\Rightarrow -\mu (R_{0s} - 1)\Psi \left[ \frac{\chi_s \chi_m (R_{0s} - R_{0m})}{R_{0s}} + \frac{\rho \omega_s \chi_m R_{0m}}{R_{0s}} \right] < 0$,

$\Rightarrow \mu (R_{0s} - 1)\Psi \frac{\chi_s \chi_m (R_{0s} - R_{0m})}{R_{0s}} \left[ 1 + \frac{\rho \omega_s R_{0m}}{\chi_s (R_{0s} - R_{0m})} \right] > 0$,

$\Rightarrow \mu (R_{0s} - 1) \frac{\chi_s \chi_m (R_{0s} - R_{0m})}{R_{0s}} \frac{\Psi}{\Psi} > 0$,

$\Rightarrow \frac{\mu \chi_s \chi_m}{R_{0s}} (R_{0s} - 1)(R_{0s} - R_{0m}) > 0$, which is true if $R_{0s} > 1$ and $R_{0s} > R_{0m}$. In the fifth line we have substituted in the definition of $\Psi$ given in equation (14). Finally if we multiply the expressions for $A_1$ and $A_2$ it is straightforward to show that the condition $A_1 A_2 > A_3$ is satisfied. Thus, by the Routh-Hurwitz criterion, the co-existent endemic equilibrium $E^2$ is locally asymptotically stable when $R_{0s} > 1$ and $R_{0s} > R_{0m}$.

## 4. Sensitivity analysis

Recognizing the relative importance of the various risk factors responsible for transmission of infectious diseases is essential. The progression of the drug-resistant strain and its incidence and prevalence must be understood in order to determine how best to decrease disease burden. As demonstrated in the previous sections, the scale and severity of disease transmission is directly associated with the basic reproduction numbers $R_{0s}$ and $R_{0m}$. Here, we estimate the sensitivity indices of the reproduction numbers $R_{0s}$ and $R_{0m}$ to the model parameters given in Table 1. The indices express how vital each parameter is to $R_{0s}$ and $R_{0m}$, and, in turn, disease transmission thus allowing us to identify which parameters should be targeted by intervention policies.

For this purpose, we calculated the partial rank correlation coefficient (PRCC) which is a global sensitivity analysis technique using Latin Hypercube Sampling (LHS). Specifically, a uniform distribution is assigned to each parameter and a total of 100,000,000 simulations are implemented. Here the model outputs we consider are both basic reproduction numbers, namely $R_{0s}$ and $R_{0m}$ as well as the sum of infectious individuals $I_s$ and $I_m$ and their total sum $(I_s + I_m)$ at equilibrium. Note that the PRCC values lie between -1 and +1. Positive (negative) values imply a positive (negative) correlation to the model parameter and outcomes. The bigger (smaller) the absolute value of the PRCC, the greater (lesser) the correlation of the parameter to the model outcome.

Fig. 2 and Fig. 3 depict the correlation between $R_{0s}$, $R_{0m}$, and the corresponding model parameters. Parameters $\beta_s$ and $\beta_m$ have positive PRCC values, implying that a positive change of these parameters will increase the basic reproduction numbers $R_{0s}$ and $R_{0m}$ respectively. In contrast, parameters $\omega_s$, and $\phi_s$ as well as $\omega_m$, and $\phi_m$ have negative PRCC values, which implies that raising these parameters will consequently decrease $R_{0s}$ and $R_{0m}$, respectively.

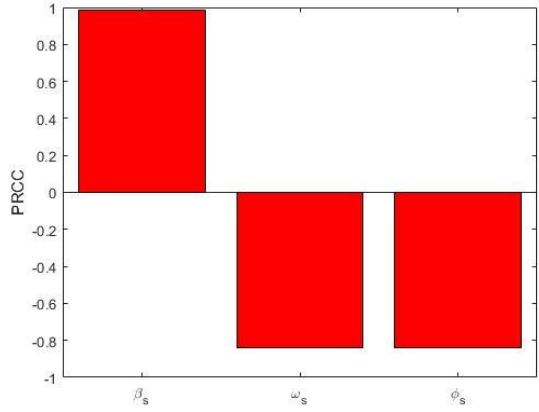 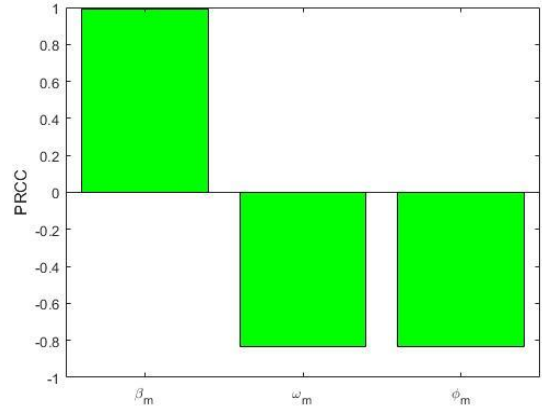

Fig. 2. PRCC values depicting the sensitivities of the model output $R_{0s}$ with respect to the estimated parameters $\beta_s$, $\omega_s$, and $\phi_s$.

Fig. 3. PRCC values depicting the sensitivities of the model output $R_{0m}$ with respect to the estimated parameters $\beta_m$, $\omega_m$, and $\phi_m$.

Fig. 4, Fig. 5, and Fig. 6 display the correlation between $I_s$, $I_m$ and $(I_s + I_m)$ and the corresponding parameters $\beta_s, \omega_s, \phi_s, \beta_m, \omega_m, \phi_m$ and $\rho$ when $R_{0s} > \max[R_{0m}, 1]$. From Fig. 4, Fig. 5, and Fig. 6, it is easy to perceive that $I_s$ and $(I_s + I_m)$ has a strong positive correlation with $\beta_s$ and $I_m$ has a weaker positive correlation with $\beta_s$, implying that a positive change of $\beta_s$ will increase $I_s$, $(I_s + I_m)$ and $I_m$. Parameters $\omega_s$ and $\phi_s$ have a negative correlation with $I_s$, $I_m$ and $(I_s + I_m)$. In addition $\beta_m$ has a negative correlation with $I_s$ and $(I_s + I_m)$ but a strong positive correlation with $I_m$. Parameters $\omega_m$ and $\phi_m$ have a positive correlation with $I_s$ and $(I_s + I_m)$ but strong negative correlation with $I_m$.

Further, parameter $\rho$ has a negative correlation with $I_s$ and $(I_s + I_m)$ but a strong positive correlation with $I_m$. Fig. 7 represents the correlation between equilibrium value of $I_m$ and the corresponding model parameters $\beta_s, \omega_s, \phi_s, \beta_m, \omega_m, \phi_m$ and $\rho$ when $R_{0m} > R_{0s}$ and $R_{0m} > 1$. Parameters $\beta_s, \beta_m$ and $\rho$ (small value not showing) have positive PRCC values, implying that a positive change in these parameters will increase $I_m$. In contrast, $\omega_s, \phi_s, \omega_m$ and $\phi_m$ have negative PRCC values and, thus, increasing theses parameters will consequently decrease $I_m$.

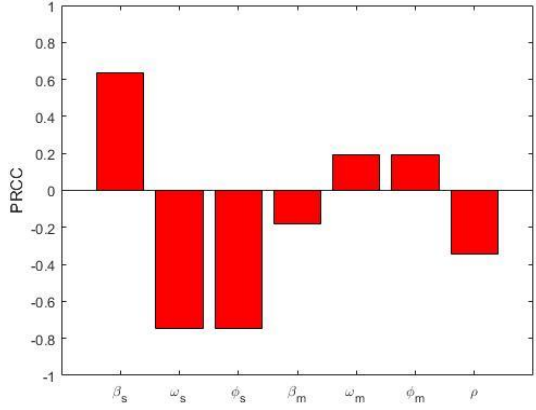

Fig. 4. PRCC values depicting the sensitivities of the model output $I_s$ with respect to the estimated parameters $\beta_s$, $\omega_s$, $\phi_s$, $\beta_m$, $\omega_m$, $\phi_m$ and $\rho$, when $R_{0s} > \max[R_{0m}, 1]$.

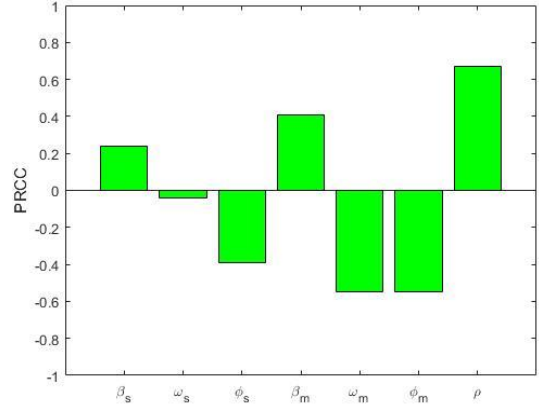

Fig. 5. PRCC values depicting the sensitivities of the model output $I_m$ with respect to the estimated parameters $\beta_s$, $\omega_s$, $\phi_s$, $\beta_m$, $\omega_m$, $\phi_m$ and $\rho$, when $R_{0s} > \max[R_{0m}, 1]$.

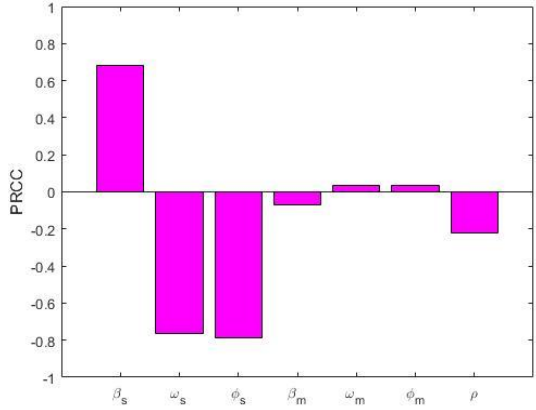

Fig. 6. PRCC values depicting the sensitivities of the model output $I_s + I_m$ with respect to the estimated parameters $\beta_s$, $\omega_s$, $\phi_s$, $\beta_m$, $\omega_m$, $\phi_m$ and $\rho$, $R_{0s} > \max[R_{0m}, 1]$.

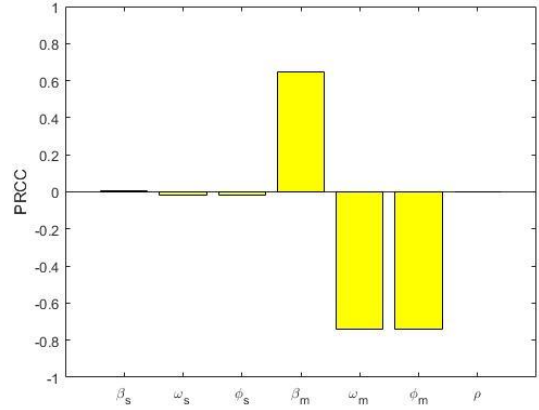

Fig. 7. PRCC values depicting the sensitivities of the model output $I_m$ with respect to the estimated parameters $\beta_s$, $\omega_s$, $\phi_s$, $\beta_m$, $\omega_m$, $\phi_m$ and $\rho$, when $R_{0m} > R_{0s}$ and $R_{0m} > 1$.

From the explicit expressions for $R_{0s}$ and $R_{0m}$ given in the equation (a)-(b), analytical expressions for the sensitivity indices can be derived following the method in [23]. For example, for a model parameter $\delta$ we would have

$$\Upsilon_{\beta_s}^{R_{0s}} = \frac{\partial R_{0s}}{\partial \delta} \times \frac{\delta}{R_{0s}}.$$

This sensitivity index is a local measure of model behaviour in terms of the model inputs. Now using the parameter values in Table 1, we have the following results (Table 2).

**Table 2:** Sensitivity indices to parameters for the model (2)-(5)

| Parameter | Sensitivity index ($R_{0s}$) | Parameter | Sensitivity index ($R_{0m}$) |
|---|---|---|---|
| $\beta_s$ | + 1.000 | $\beta_m$ | +1.000 |
| $\omega_s$ | - 0.431 | $\omega_m$ | - 0.274 |
| $\phi_s$ | - 0.548 | $\phi_m$ | - 0.699 |

In the sensitivity indices of $R_{0s}$ and $R_{0m}$, the most sensitive parameters are the effective contact rate of drug-susceptible case, $\beta_s$ and drug-resistant case, $\beta_m$, respectively. Other significant parameters are recovery rates ($\omega_s$ and $\omega_m$) and disease related death rates ($\phi_s$ and $\phi_s$). Since $\Upsilon_{\beta_s}^{R_{0s}} = 1$, and $\Upsilon_{\beta_m}^{R_{0m}} = 1$, increasing (or decreasing) the effective contact rates, $\beta_s$ and $\beta_m$ of drug-susceptible and drug-resistant cases by 100%, increases (or decreases) the reproduction numbers $R_{0s}$ and $R_{0m}$ by 100%. Similarly, increasing (or decreasing) the recovery rates $\omega_s$ and $\omega_m$ by 100% decreases (or increases) $R_{0s}$ and $R_{0m}$, by 43.1% and 27.4% respectively. In the same manner, decreasing (or increasing) the disease related death rates $\phi_s$ and $\phi_m$ by 100% increases (or decreases) the $R_{0s}$ and $R_{0m}$, by 54.8% and 69.9% respectively.

## 5. Numerical simulations

In this section, we carry out detailed numerical simulations (using the Matlab programming language) to support the analytic results and to assess the impact of amplification and drug-susceptible treatment/recovery rate on equilibrium levels of total prevalence and drug-resistant prevalence. Three equilibrium points were found: the disease-free equilibrium $E^0$; a mono-existent endemic equilibrium $E^1$; and co-existent endemic equilibrium $E^2$. Routh-Hurwitz conditions and the Lyapunov function were used to investigate the local and global stability of these points. We used different initial conditions for both strains of all populations and found that if both basic reproduction numbers are less than one (i.e. $\max[R_{0s}, R_{0m}] < 1$) then the disease free equilibrium is locally and globally asymptotically stable. If $R_{0m} > \max[R_{0s}, 1]$, and the drug-susceptible strain dies out but the drug-resistant strain persists in the population. Furthermore, if $R_{0s} > \max[R_{0m}, 1]$, then both the drug-susceptible and drug-resistant strains persist in the population. Fig. 8 represents co-existent endemic equilibrium and we used different initial conditions for this system trajectories in the $I_s$ vs $I_m$ plane. In this system both strains ($I_s$ and $I_m$) persisting; this is because the basic reproduction number of the drug-susceptible strain $R_{0s}$ was greater than one and there was an amplification pathway from the drug-susceptible strain to the drug-resistant strain (i.e. $R_{0s} > \max[R_{0m}, 1]$).

Fig. 9 depicts the effect of amplification (ρ) on equilibrium levels of drug-susceptible prevalence and drug-resistant prevalence and shows that in the first region (ρ ≲ 0.6) the drug-susceptible prevalence is initially dominant but with the drug-resistant prevalence rising with increasing ρ. Eventually, for ρ ≳ 0.6, the drug-resistant strain becomes dominant courtesy of the amplification pathway. Fig. 10, and Fig. 11 are graphical representations showing the effect of drug-susceptible strain treatment/recovery rate on the equilibrium level of total prevalence, and drug-resistant prevalence when both infection rates ($\beta_s$, $\beta_m$) are fixed. If we increase the proportion of amplification, both the total prevalence and drug-resistant prevalence also increase. However, Fig. 11 shows that for high amplification, the drug-resistant prevalence increased when the treatment/recovery rate of the drug-susceptible strain moved from zero to around 0.25 to 0.30, then declined to a common point. For lower amplification values, the drug-resistant proportion only increased up to the common point. This point is the drug-resistant-only equilibrium and occurs when the effective reproduction number of drug-susceptible strain becomes lower than the basic reproduction ratio of drug-resistant strain. Numerical simulations show that for sufficiently high amplification, the prevalence of the drug-resistant strain will exceed that of its inherent equilibrium value (that is, the resistant-only equilibrium) when the drug-susceptible strain exists and is being treated.

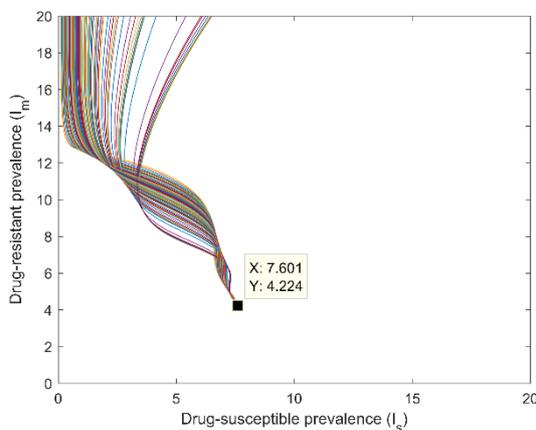

Fig. 8. Co-existent endemic equilibrium: $R_{0s} > \max[R_{0m}, 1]$. In this case both drug-susceptible prevalence and drug-resistant prevalence persist in the population.

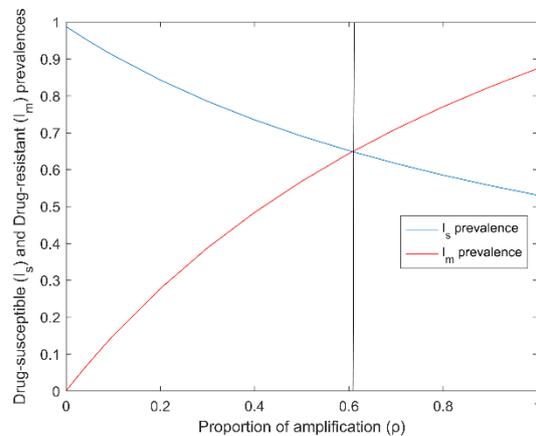

Fig. 9. Effect of amplification (ρ) on the drug-susceptible ($I_s$) prevalence and drug-resistant prevalence ($I_m$). All remaining parameter values assume their baseline values given in Table 1.

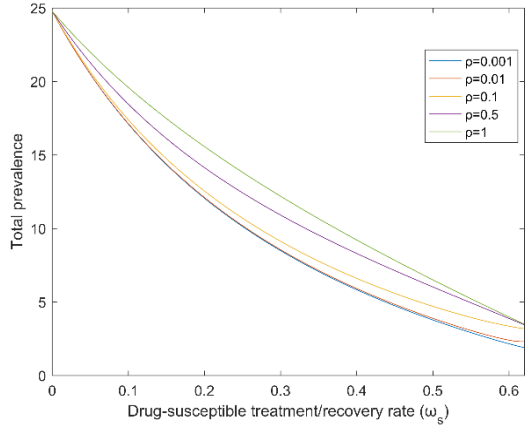

Fig. 10. Effect of drug-susceptible treatment/recovery rate ($\omega_s$) on equilibrium level of total prevalence when both infectious rates ($\beta_s$, $\beta_m$) are fixed. All remaining parameter values assume their baseline values given in Table 1.

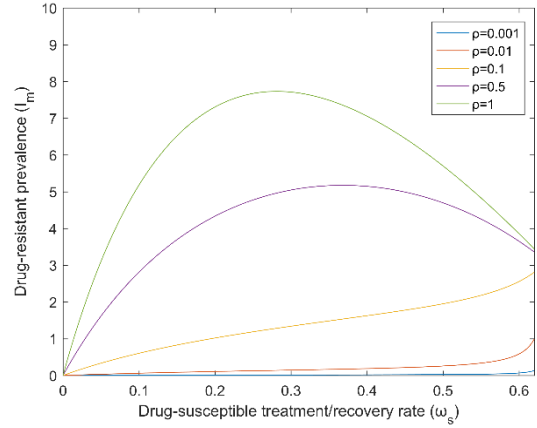

Fig. 11. Effect of drug-susceptible treatment/recovery rate ($\omega_s$) on equilibrium level of drug-resistant strain when both infectious rates ($\beta_s$, $\beta_m$) are fixed. All remaining parameter values assume their baseline values given in Table 1.

## 6. Discussion and conclusion

In this study, we formulated a two-strain SIR non-constant population model with amplification and investigated its dynamic behavior. We considered amplification as the process by which an individual infected with a drug-susceptible strain acquires infection with a drug-resistant strain. Using the next generation matrix, we obtained the basic reproduction number of each strain, namely $R_{0s}$ for drug-susceptible cases and $R_{0m}$ for drug-resistant cases. We found that the basic reproduction numbers determine the equilibrium states of the system and their stability. Specifically if $R_{0m}$ is greater than $R_{0s}$ and unity, only the drug resistant strain will remain, whereas if $R_{0s}$ is larger than $R_{0m}$ and unity, a coexistence is likely. We also found that both basic reproduction numbers are independent of amplification rates, which indicates that the reproductive capacity of each strain is autonomous of the amplification rates between them.

We then determined the necessary and sufficient conditions for disease extinction and disease endemicity according to the values of the basic reproduction numbers. As expected, the infected population of both strains declines rapidly and is extinguished if both basic reproduction numbers are less than one ($\max[R_{0s}, R_{0m}] < 1$). Conversely, if $R_{0m} > \max[R_{0s}, 1]$, then the drug-susceptible strain dies out but the drug-resistant strain persists in the population. Furthermore, if $R_{0s} > \max[R_{0m}, 1]$, then both the drug-susceptible and drug-resistant strains persist in the population.

We also found that the drug-susceptible strain is not necessarily the most prevalent at equilibrium even if it has the highest basic reproduction number. This is a consequence of the fact that the drug-susceptible strain persists purely on direct transmission whereas the drug-resistant strain prevalence is driven by a combination of direct transmission and amplification. These results explain in part the rise in drug-resistant strain prevalence when the drug-susceptible strain is treated. These findings can hopefully advise public health policy makers to formulate policy for drug-resistant prevalence since we have shown that the drug-resistant strain can surpass the drug-susceptible strain (i.e. become more prevalent) even if drug-resistant strain has a smaller basic reproduction number.

Lastly we explored the effect of the drug-susceptible treatment rate on the equilibrium level of total prevalence and drug-resistant prevalence. We found that if we increase the drug-susceptible treatment rate, the total prevalence will decline. However, the response of the drug-resistant strain prevalence is non-monotonic, increasing for a certain period and then declining at a particular threshold point. This finding has important implications for choosing the proper intervention or treatment strategies. From a microbiological viewpoint, resistance first occurs by a genetic mutation in a micro-organism that leads to resistance to a treatment, modelled by reducing the treatment rate. Therefore, one could question whether it is prudent to risk the emergence of drug resistant strains by increasing the treatment rate of the drug-susceptible strain. However, at least initially, such resistance-conferring mutations typically exact a "fitness cost" whereby drug-resistant organisms reproduce at a lower rate and are often less transmissible than their drug-susceptible counterparts [24]. However, the selective pressure applied by antibiotic treatment permits drug-resistant mutants to become the dominant strain in a patient infected with disease on first-line therapy and allows for further mutations with selection for fitness. Therefore, increasing drug-susceptible treatment rates may increase the likelihood of emergence of an even more prolific strain which also has drug resistance.

In conclusion, this study has concentrated mainly on a two-strain coupled SIR epidemic model and performed a rigorous analytical analysis of the system properties and solutions, for understanding infectious diseases genetic variation and the rising threat of antibiotic resistance or inadequate treatment. These results help inform the practice of drug treatment in the setting of drug resistance and emergent strains, such as is occurring in tuberculosis and other bacterial pathogens. Future studies could focus on specific pathogens (and their associated parameters) and whether treatment may lead to unintended threats to infection control such as increase in resistant strains.

**Acknowledgements**

The authors would like to thank Dr. Elizabeth Tynan, James Cook University, for her assistance in preparing this article.